\documentclass[aps,twocolumn,secnumarabic,nobalancelastpage,amsmath,amssymb,
nofootinbib]{revtex4}

\usepackage[T1]{fontenc}
\usepackage[latin9]{inputenc}
\usepackage{textcomp}
\usepackage{amsmath}
\usepackage{amssymb}
\usepackage{esint}
\usepackage{graphics}      
\usepackage{graphicx}      
\usepackage{longtable}     
\usepackage{url}           
\usepackage{bm}            
\usepackage{color}

\begin{document}
\title{From Compton Scattering of photons on targets to              
                                                             
Inverse Compton Scattering of electron and photon beams}
\author         {Luca Serafini, Vittoria Petrillo }

\affiliation    {INFN Sez. Milano and Universita' degli Studi di Milano, Dipartimento di Fisica, Via Celoria 16, 20133, Milano (Italy)}
\date{\today}

\begin{abstract}

We revisit the kinematics of Compton Scattering (electron-photon interactions producing electrons and photons in the exit channel) covering the full range of energy/momenta distribution between the two colliding particles, with a dedicated view to statistical properties of secondary beams that are generated in beam-beam collisions. Starting from the Thomson inverse scattering, where electrons do not recoil and photons are back-scattered to higher energies by a Lorentz boost effect (factor $4\gamma^2$), we analyze three transition points, separating four regions. These are in sequence, given by increasing the electron recoil (numbers are for transition points, letters for regions): a) Thomson back-scattering, 1) equal sharing of total energy in the exit channel between electron and photon, b) deep recoil regime where the bandwidth/energy spread of the two interacting beams are exchanged in the exit channel, 2) electron is stopped, i.e. taken down at rest in the laboratory system by colliding with an incident photon of $mc^2/2$ energy, c) electron back-scattering region, where incident electron is back-scattered by the incident photon, 3) symmetric scattering, when the incident particles carry equal and opposite momenta, so that in the exit channel they are back-scattered with same energy/momenta, d) Compton scattering ($a'$ $la$ Arthur Compton, see ref.4), where photons carry an energy much larger than the colliding electron energy. For each region and/or transition point we discuss the potential effects of interest in diverse areas, like generating mono-chromatic gamma ray beams in deep recoil regions with spectral purification, or possible mechanisms of generation and propagation of very high energy photons in the cosmological domain.
\end{abstract}

\maketitle

\section{Introduction}

After the formulation of special relativity theory  and the derivation of the relativistic Doppler effect \cite{Einstein}, predicting a blu-shift of e.m. field frequency seen by an observer in relativistic motion counter-propagating with respect to the direction of propagation of the electro-magnetic wave (see Ref. \cite{Thomson}), confirmed by experimental measurements (\cite{Stilwell}), the theoretical frame of Classical Electro-Dynamics was ready to explain and quantitatively calculate the characteristics of radiation generated by charged particles moving in space at relativistic velocities. Such a theoretical frame has been made available in the early '900 by A. Einstein, J.J. Thomson and J. Larmor, to predict a broad class of phenomena of radiation emission mechanisms where the quantum nature of e.m. fields was not needed nor critical. Only after A. Compton's fundamental work \cite{Compton} the quantum nature of e.m. fields, i.e. the photon, was experimentally demonstrated: its existence was a game changer and basically showed through A.Compton's measurements that the effect of electron recoil produced by the X-ray photon in the collision of X-rays with atomic electrons couldn't be explained by classical electro-dynamics and special relativity. The red-shift of back-scattered photons could be explained only invoking a photon-electron collision where the electron recoils and takes out some of the photon energy in form of kinetic energy. After this discovery the electron recoil became the fundamental parameter, i.e. the continental divide, between the classical picture of radiation emission and the quantum model of photon emission by charged particles propagating in space at relativistic velocities. In this paper we revisit the full range of values that the recoil parameter can assume, from the very small values associated to Thomson scattering (or Thomson back-scattering depending on the reference frame where the electron is observed) up to the deep recoil region, defined as when the incident photon has an energy much larger than $mc^2$ in the electron rest frame (step A. Compton!), underlying four different domains and three transition points separating them. 
The recoil factor $X$ is a dimensionless parameter given by :
\begin{equation}
 X = 4E_{ph}E_e/(mc^2)^2 = 4 \gamma E_{ph}/mc^2 = 4\gamma^2E_{ph}/E_e   
\end{equation}
where $E_e$ is the (total) electron energy and $E_{ph}$ is the photon energy before the interaction, and the electron Lorentz factor gamma is given by $\gamma = E_e/mc^2$. The recoil factor $X$ is linked to the energy available in the center of mass reference system $E_{cm} $ by $E_{cm}=mc^2\sqrt{(1+X)}$.
The scope of this paper is to underline the specific behaviors and statistical properties of photon beams generated in correspondence of such transition points and regions. Several applications and investigation areas can benefit from these properties: we anticipate here the spectral purification mechanism active in the deep recoil region, the surprising property of X-rays with energy equal to $mc^2/2$ to stop any colliding relativistic electron of any arbitrarily high kinetic energy, and the cancellation of the $\gamma^2\theta^2$ angular correlation, inherent in any relativistic Doppler based radiation source, occurring at the Symmetric Compton Scattering transition point \cite{Serafini}. Historically, the progress in understanding Compton scattering in its whole dynamical range, following the fundamental work of A. Compton, who studied the phenomenon in the frame of direct scattering of energetic (X-ray) photons interacting with electrons at rest in the laboratory frame, has been initially pursued by researchers active in astrophysics, trying to explain the behaviour of high energy charged particles propagating throughout the universe, interacting with photons of the cosmic back-ground (microwave, infrared, optical). J.W. Follin in 1947 firstly studied, in the course of his PhD thesis project \cite{Follin}, the inverse kinematics of a relativistic electron colliding with a low energy (< 1 eV) photon where the term inverse defines a specular mode of interaction w.r.t A.Compton's one, i.e. a configuration in the interaction where the photon gains energy in the collision at expense of the electron kinetic energy, the opposite of what happens in direct Compton scattering. Follin's study has been re-issued one year later by Feenberg and Primakoff \cite{Primakoff} still in the frame of astrophysical phenomena, where inverse Compton scattering plays a crucial role in determining the universe opacity to high energy photons and electrons \footnote{Previous literature on inverse Compton attributed to ref.  \cite{Primakoff} the first interpretation of inverse Compton - we discovered from reference 2 of Feenberg's paper that Follin developed and published his study one year earlier - Feenberg and his coauthor Primakoff were prompted by J.R. Oppenheimer to read Follin's work that anticipated their study}. Therefore, the first studies of Inverse Compton Scattering (ICS)  dated back to the late '40s and had the aim to explain astrophysical phenomena related to the first observations of cosmic rays carried out just after W.W.II. It's been only fifteen years later, with the advent of high energy (GeV-class) electron synchrotrons and the invention of the ruby-laser,  that R.H. Milburn \cite{Milburn} firstly conceived the possibility to perform a laboratory-based experiment of ICS using the relativistic electron beams generated by particle accelerators making them collide with the intense optical photon beams made available by lasers. Milburn was quickly followed by Arutyunian and Tumanian's theoretical analysis \cite{Arutyunian}. The crucial role of recoil factor $X$ was fully recognized since the initial stages of these theoretical studies, both in the astrophysical domain and in the accelerator-laser based scenario: while in the astrophysical domain Follin and Feenberg classified just two main regimes of the electron-photon collision kinematics, i.e. the low relativistic regime $X\ll1$ and the high relativistic regime $X\gg 1$ , both accessible in the cosmic ray contest, Milburn and Arutyunian focused mostly on the low to mild relativistic regime ($X<1$) that was only accessible in laboratory experiments. The first measurements of ICS at a laboratory scale followed a few years after Milburn's publication \cite{ Telnov, Bemporad, Ladon} at a number of particle accelerator laboratories. We organize this paper as follows: in Section 2 we discuss the full range of inverse kinematics reporting a synoptic table listing all ICS regions and transition points. In Section 3 we show how the dependence on the collision angle is also vanishing in large recoil regimes, and in Section 4 we show some quantitative examples of spectral purification, the relevant aspect of large recoil electron-photon collisions, where the properties of the incident primary beams are mapped into the secondary beams with an exchange of entropy, i.e. the scattered photon beam is cooled and the exit electron beam after collision is heated, a strategic way to generate mono-chromatic gamma rays using mono-energetic incident electron beams and broad-band white-spectrum photon beams We should remark that in this paper we are not discussing non-linear effects due to intensity of the incident photon beam, i.e. those effects who become important when the laser parameter $a_0$ associated to the optical photon beam carried by an incident laser pulse becomes close or larger than 1, inducing multi-photon or dressed-electron effects.

\section{Different Compton Scattering regimes}

In the synoptic table \ref{tab:1} the different Compton Scattering regimes are summarized. Three transition points are listed in the table rows, separating four regions, each characterized by specific range of collision kinematics. 
\begin{table*}
\caption{\label{tab:1} Regimes of Compton Scattering. }

\begin{tabular}{|c|c|c|c|c|c|c|c|}
\hline 
Regime & Recoil Factor  & Inc. ph. en.  & Max scatt. ph. en.  & Min scatt. e-en.&Cent. of mass en. & El. rest frame \tabularnewline
\hline 
 Symbol & $X$ & $E_{ph}$ & $E'_{ph}$  & $E'_e$ & $E_{cm}$ & ${E_{ph}}^{ERF}$  \tabularnewline
\hline
Definition & $X=\frac{4\gamma E_{ph}}{mc^2}$ & $k_x=k_y =0 $ &$E'_{ph}=\frac{4\gamma^2 E_{ph}}{1+X}$ &&$E_{cm}=mc^2\sqrt{1+X}$ & $2\gamma E_{ph}$  \tabularnewline
\hline

ITS &$X \ll 1$ & $ \ll E_e$ &$4\gamma^2E{ph}$ &$\simeq E_e$&$\simeq mc^2$  &$<<mc^2$ \tabularnewline
\hline 
DICS & X=1 & $\frac{mc^2}{4\gamma}$ &$2\gamma^2 E_{ph}$ &$2\gamma^2 E_{ph}$&$\sqrt{2}mc^2$ & $mc^2/2$\tabularnewline
\hline 
DRCS& $1\ll X \ll 2\gamma$ & $ \gg \frac{mc^2}{4\gamma}$ &$\simeq(1-\frac{1}{X})E_e$ & $ \ll E_e$& $>\sqrt{2}mc^2$&>$mc^2/2$ \tabularnewline
\hline 

FICS &$X=2\gamma$ & $mc^2/2$ &$E_e-mc^2/2$ &$mc^2$& $\sqrt{2\gamma}mc^2$& $\gamma mc^2$\tabularnewline
\hline 
 EBS& $2\gamma<X<4\gamma^2$& <$\gamma mc^2$ &$\sim E_e$ & $<E_e$ &>$\sqrt{2\gamma}mc^2$& >$\gamma mc^2$\tabularnewline
\hline 
SYCS & $X=4\gamma^2$&  $\gamma mc^2$&$E_{ph}$ &$E_e$ &$2\gamma mc^2$&$2\gamma^2mc^2$ \tabularnewline
\hline 
RDCS & $X>4\gamma^2$& $>\gamma mc^2$ &$<E_{ph}$ & $>E_e$&>$2\gamma mc^2$&>$2\gamma^2mc^2$ \tabularnewline
\hline

\end{tabular}

\end{table*}
Table columns report all relevant parameters characterizing the electron-photon collision quantities: recoil factor $X$, energy of incident photon $E_{ph}$, maximum energy $E'_{ph}$ of scattered photon at $\theta=0$ scattering angle (full back-scattering), the corresponding minimum energy $E'_e$ of the scattered electron, the energy in the center of mass reference frame $E_{cm}$ and the energy of the incident photon as seen in the electron rest frame $E_{ph}^{ERF}$ . Note that the incident electron is assumed to be relativistic (i.e.  $\gamma \gg 1$) and the electron-photon collision is assumed to be head-on (i.e. vanishing transverse momentum of the incident photon and electron). 
 Besides the recoil factor $X$, a second dimensionless parameter governing the kinematics of Compton scattering is the asymmetry parameter $A$, defined as in ref. \cite{Serafini}
 
 \begin{equation}
 A = \gamma^2  -X/4    
 \end{equation} 
 (in case of relativistic electrons). 
 $A$ represents the divide between direct Compton scattering (characterized by $A<0$, where the energy of the incident photon is larger than the energy of the incoming electron), and inverse Compton scattering (region defined by $A>0$) where the electron energy before interaction is larger than that of the incident photon. $A=0$ is the transition point where the scattering is fully symmetric (actually the center of mass reference system is steady in the  laboratory system because the total momentum is null) and was denominated Symmetric Compton Scattering in Ref. \cite{Serafini}.

For simplicity in the discussion we will consider in this section an ideal head-on collision between an electron of energy $E_e$ and momentum aligned with the z-axis, directed towards positive z, and a photon of energy $E_{ph}$, counter-propagating w.r.t. the electron towards negative z. The generalization to an arbitrary value of the collision angle $\alpha$ between the incoming electron and the incident photon will be given in Section 3. The electron momentum before collision is $\textbf{P} = (p_x=0,p_y=0,p_z=\beta\gamma mc)$ while the photon momentum is $\hbar \textbf{k} = (\hbar k_x = 0, \hbar k_y = 0,\hbar k_z = -E_{ph}/c)$. Following references \cite{Ranjan,Curatolo} and invoking the conservation of total energy and momentum, if the scattered photon is propagating at a scattering angle $\theta$ w.r.t. the z-axis, its energy will be given by the well known formula:
\begin {equation}
E'_{ph}=\frac{(1+\beta)E_{ph}E_e}{(1-\beta cos\theta)E_e+(1+cos\theta)E_{ph}}
\end{equation}

that is valid for any arbitrary value of colliding electron and photon energy and momentum. Actually, the direct Compton effect is represented by Eq. 3 just setting $\beta=0$ and $E_e=mc^2$ , that applies to an electron at rest in the laboratory system. In this way, Eq.3 transforms into the well know Compton's formula : 
\begin{equation}
E'_{ph}=\frac{E_{ph}}{1+(1+cos\theta)E_{ph}/mc^2}
\end{equation}
By knowing the value of the scattered photon energy $E'_{ph}$, and calling $E_{tot}=E_e+E_{ph}$, it is straightforward to derive the electron energy after scattering as $E'_e = E_e + E_{ph} - E'_{ph}$ = $E_{tot} - E'_{ph}$ . Nevertheless, our focus is on relativistic electrons, so Table \ref{tab:1} illustrates the behavior of electron and photon energy after scattering in case $\gamma>>1$. 
\begin{figure*}\centering

	\includegraphics[width=10 cm]{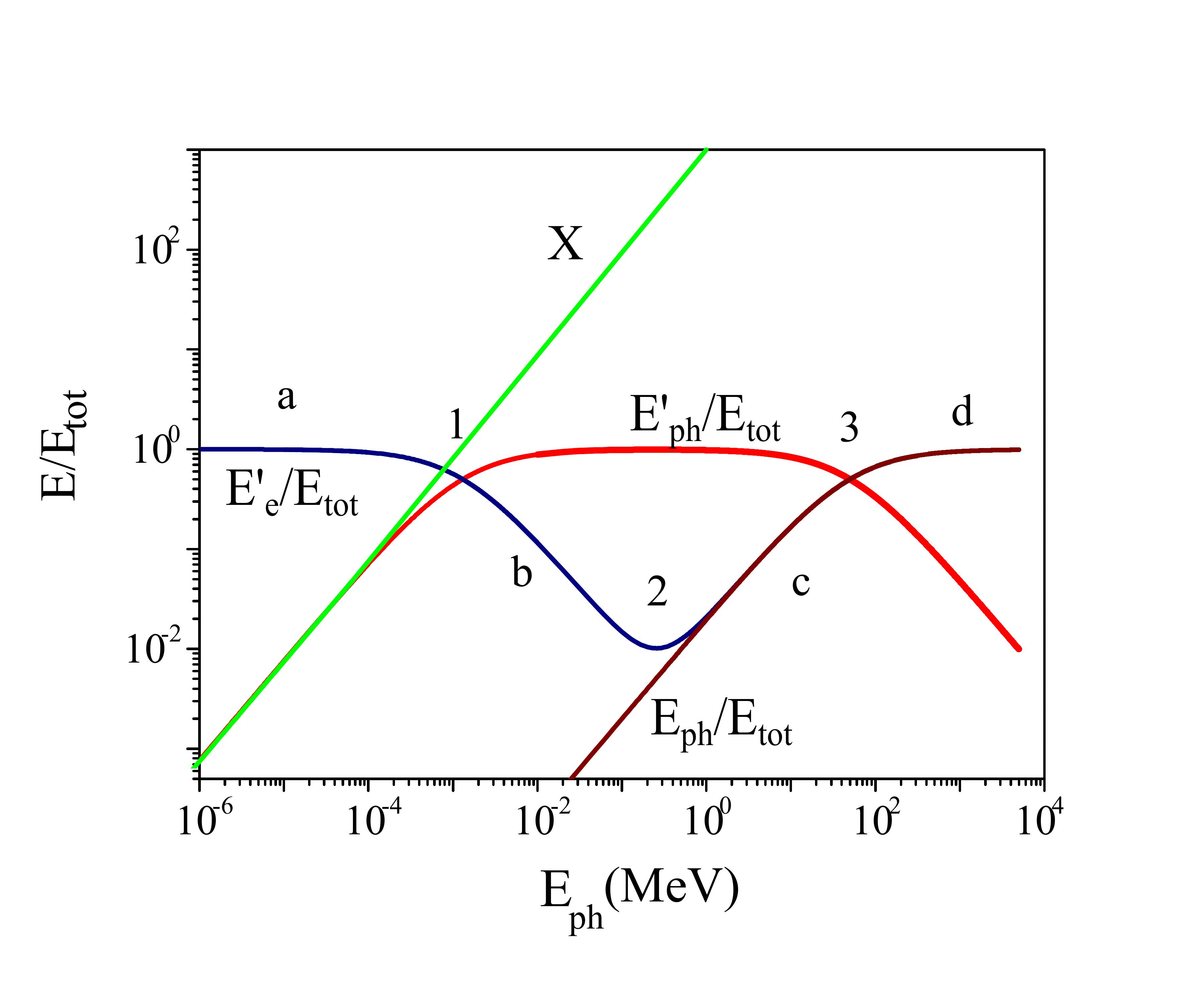}
	\caption{Different regimes of emission versus the initial photon energy (in MeV) for $E_e=50 MeV$. All energies (namely $E'_{e}$ (in blue),$ E'_{ph}$ (in red) and $E_{ph}$ (in black) are dimensionless, normalized to $E_{tot}$. The recoil $X$ (on the same left ordinate) is in green. (a) indicates the regime ITS, (b) indicates DRCS, (c) EBS and (d) RDCS; (1) represents the point DICS ,(2) represents the point FICS and (3) SYCS }
	\label{fig:regimi}
\end{figure*}
Therefore the fourth region named RDCS refers to Direct Compton effect with electrons in relativistic motion in the laboratory system, though of energy smaller than that of incident photons. In order to better illustrate the 4 regions and 3 transition points of generalized Compton scattering, we specialize in the following the discussion considering those photons that are totally back-scattered and propagate at a scattering angle $\theta=0$ after colliding with the electron. These are photons achieving maximum energy gain or loss depending on the value of the Asymmetry parameter $A$ (photons gain energy when $A>0$ and loose energy when $A<0$). If $\theta=0 $, the energy of the back-scattered photon is simply given by $E'_{ph} = 4\gamma^2 E_{ph}/(1+X)$. 
We plot in Fig. \ref{fig:regimi} the relevant quantities characterizing the kinematics as a function of the incident photon energy $E_{ph}$ for a fixed incident electron energy, namely $E_e$ = 50 MeV. All energies are normalized to the total energy of the system $E_{tot} = E_e + E_{ph}$. The recoil parameter $X$ is also plotted. The normalized energies of the electron and photon after scattering can be expressed in terms of only the recoil parameter $X$ and the Lorentz factor $\gamma$ as follows:
\begin{equation}
    \frac{E'_{ph}}{E_{tot}}=\frac{X}{(1+X)(1+\frac{X}{4\gamma^2})}
\end{equation}
\begin{equation}
    \frac{E'_e}{E_{tot}}=1-\frac{X}{(1+X)(1+\frac{X}{4\gamma^2})}
\end{equation}

\begin{figure*}\centering

	\includegraphics[width=18 cm]{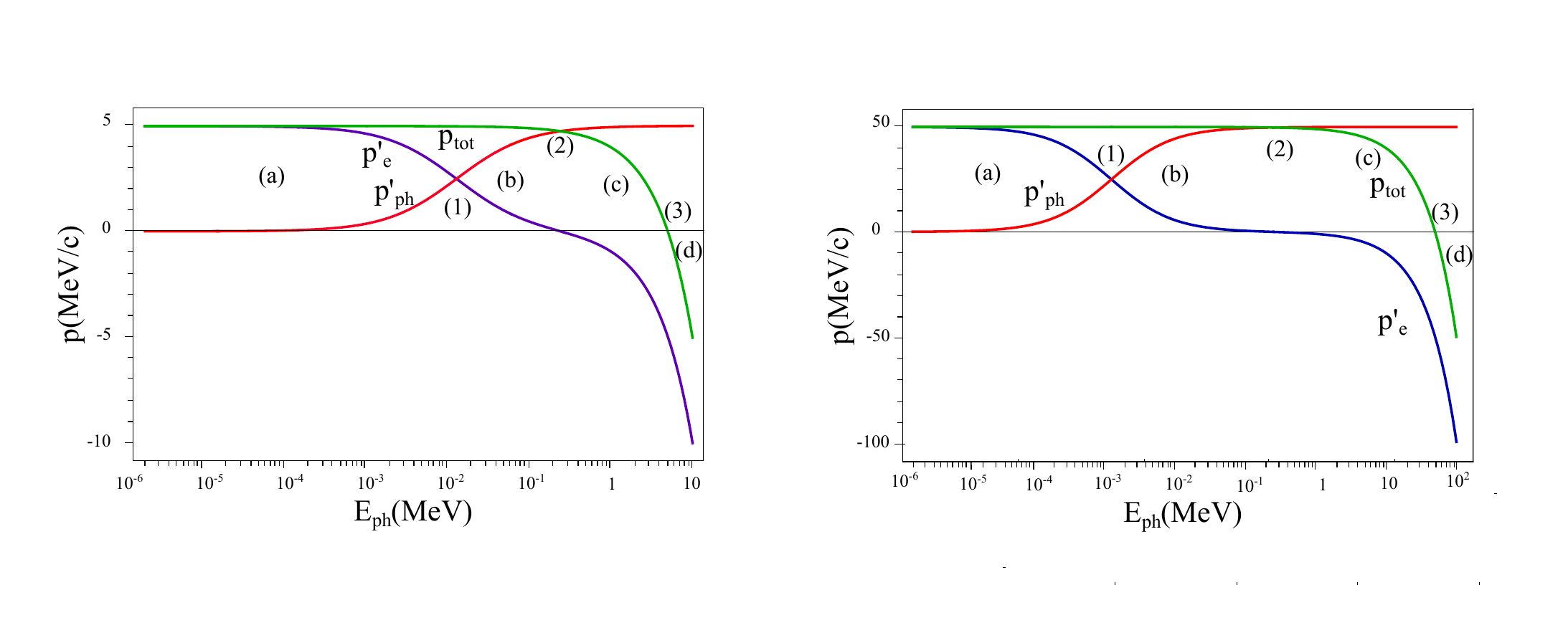}
	\caption{Momenta versus the initial photon energy (in MeV) for left $E_e=5 MeV$ and right $E_e=50 MeV$. All momenta (namely $p'_{e}$ (in blue),$ p'_{ph}$ (in red) and $p_{tot}$ (in green) are in MeV/c. (a) indicates the regime ITS, (b) indicates DRCS, (c) EBS and (d) RDCS; (1) represents the point DICS ,(2) represents the point FICS and (3) SYCS }
	\label{fig:momenti}
\end{figure*}

\begin{figure*}\centering

	\includegraphics[width=10 cm]{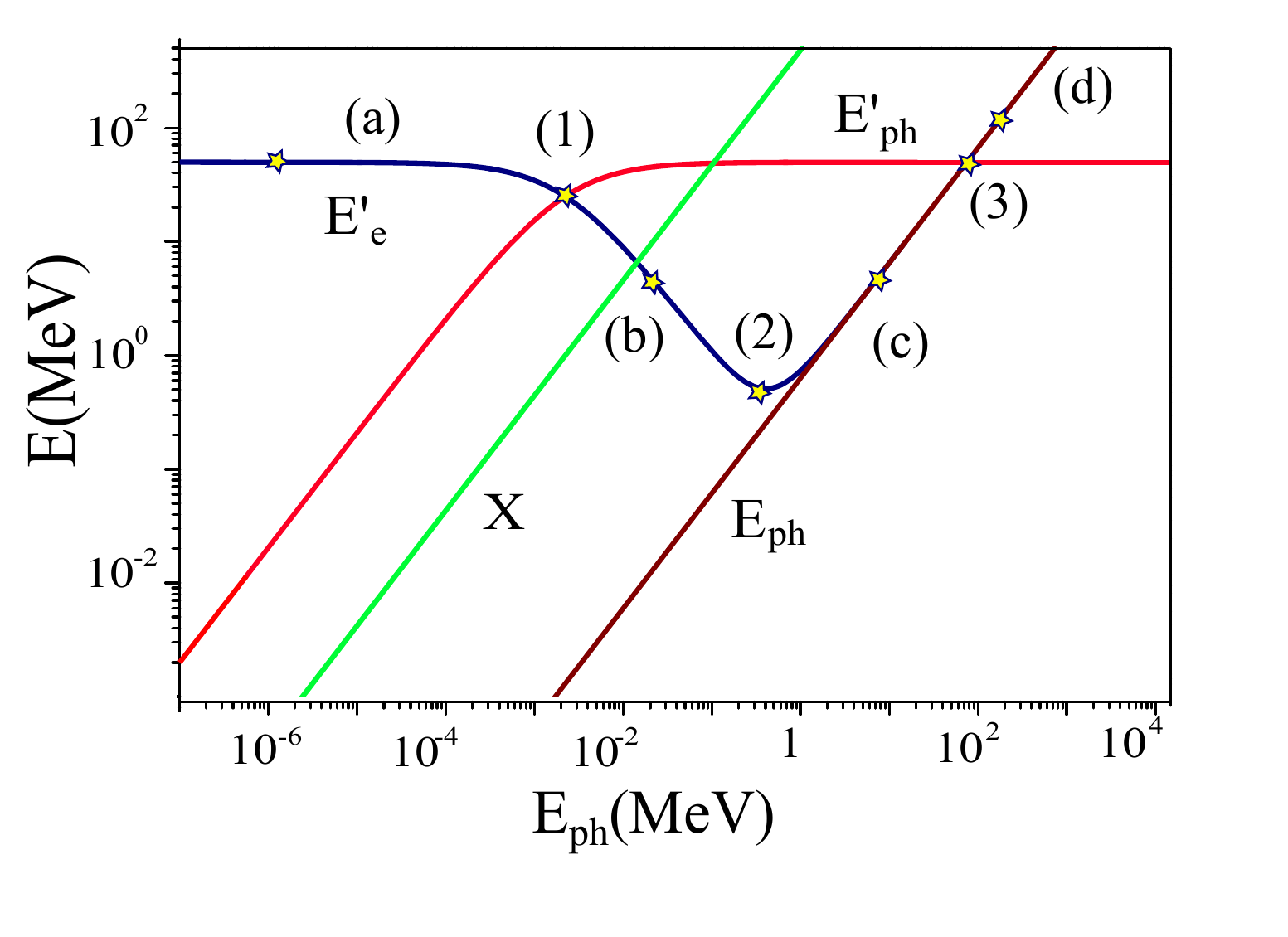}
	\caption{Different regimes of emission versus the initial photon energy (in MeV) for $E_e$= 50 MeV.  $E'_{e}$ (in blue),$ E'_{ph}$ (in red) and $E_{ph}$ (in black) in MeV are represented versus $E_{ph}$ in MeV. The recoil $X$ (on the same left ordinate) is in green. Differently from Fig. \ref{fig:regimi} the quantities (except $X$) are not dimensionless. The stars represent the values chosen for the spectrum simulations.}
	\label{fig:50_regimi}
\end{figure*}

These are very useful expressions to explain how to classify the various regimes and transition points of Compton inverse-direct scattering. 
Fig.\ref{fig:momenti} shows the total momentum $p_{tot}$, as well as the final electron and photon momenta, respectively $p'_{e}$ and $p'_{ph}$.
The first region, named Inverse Thomson Scattering (label (a) in Figs.\ref{fig:regimi} and \ref{fig:momenti}), is basically defined as that of "negligible momentum exchange" between the incident photon and the colliding electron: as a matter of fact the momentum exchange in the Electron Rest Frame (ERF) is given by $2E_{ph}^{ERF} = 4\gamma E_{ph}$ (note that $X=2E_{ph}^{ERF}/mc^2$). If the momentum exchange is negligible w.r.t. $mc^2$ then the condition $X\ll1$ holds, and the back-scattered photon in ERF will have same energy as  $E_{ph}^{ERF}$ : Lorentz transforming back to the laboratory system will give $E'_{ph}= 2\gamma E_{ph}^{ERF} = 4\gamma^2 E_{ph}$. Hence, Inverse Thomson Scattering can be simply explained as inverse photon scattering with negligible momentum exchange (note that in this regime the total energy-momentum conservation is violated by the expression $4\gamma^2E_{ph}$ unless $X\ll1$). ITS regime and its peculiar $4\gamma^2$ scaling of the back-scattered photon energy can be also explained going through a classical electro-dynamics description of undulator radiation emitted by the electron oscillating in the em. field of the counter-propagating laser pulse (see ref. \cite{Tomassini}). The majority of Inverse Compton Sources under design or operation belong to this ITS regime (see \cite{Petrillo} ), with the exception of XFELO (\cite{Hajima}). Looking at Eq. 5 and 6, it is interesting noticing that the two particles after scattering have equal energy, i.e. $E'_e=E'_{ph}$ when the recoil parameter becomes equal to either $X=1$ or $X=4\gamma^2$. We denominate the transition point characterized by $X=1$ , labelled (1) in Fig.s \ref{fig:regimi}-\ref{fig:50_regimi} , Democratic Inverse Compton Scattering, since the two particles emerging after scattering are sharing democratically the total energy of the system $E_{tot}$ and the total momentum $p_{tot}$ (while before scattering most of the energy is carried by the electron and the photon carries an energy still much smaller than $mc^2$, namely $mc^2/(4\gamma)$). The second solution $X=4\gamma^2$ , that is valid only when $\gamma\gg1$, corresponds actually to Symmetric Compton Scattering, very well discussed and analyzed in ref. \cite{Serafini}, that is characterized by interacting electron and photon carrying equal energy and momentum before and after scattering. In SYCS (point (3) in the Figures) what happens is basically a flip-over of the incoming momentum of the two scattering particles, that just flip the direction of their momentum without modifying their energy, with a uniform angular distribution after scattering.

Region (b) in Figs. \ref{fig:regimi} and \ref{fig:momenti}, denominated Deep Recoil Compton Scattering, is characterized by an increasing value of the recoil factor, in the range $1\ll X\ll 2\gamma$. 
In this region the incident electron is going to lose most of its energy/momentum in favour of the back-scattered photon: a saturation effect starts to appear in the energy of the back-scattered photon, given approximately by $(1-1/X)E_e$ , which depends only perturbatively on the energy of the incident photon (through the recoil factor $X$). This is the basis of spectral purification mechanism that is further discussed in section 4. Still in region (b) the scattered electron emerging after scattering keeps propagating towards the incidental positive z-axis direction, as clearly shown in Fig.2. Instead, in region (c), denominated Electron Back-Scattering, characterized still by a large value of the recoil factor, the electron starts to be back-scattered and propagating back in the negative direction of the z-axis after scattering. Both in region (b) and region (c), the emerging photon after scattering is carrying almost all the total energy of the system, since $E'_{ph}/E_{tot}\simeq 1$ . 

\begin{table*}
\caption{\label{tab:2} Parameters of simulations }

\begin{tabular}{|c|c|c|c|c|c|c|c|c|}
\hline 
Regime & $E_e$  & $E_ph$  & X &A&Max scatt ph. En.& Min scatt. e-En.&Cent. of mass En. &Figure\tabularnewline
\hline
Unit&MeV&MeV&&&MeV&MeV&MeV&\tabularnewline
\hline

ITS &50&1.5$\times10^{-6}$& 0.0011&9574&0.057&50&0.511&Fig.\ref{fig:ICS}\tabularnewline
\hline 
DICS &50&1.3$\times 10^{-3}$&1&9574&24.88 &24.88&0.722&Fig.\ref{fig:ICS}\tabularnewline
\hline 
DRCS& 50&$10^{-2}$&7.65&9572&43.46&6.5&1.5&Fig.\ref{fig:ICS}\tabularnewline
\hline 

FICS&50 &0.2555&195.7&9525&50&0.2555&7.1&Fig.\ref{fig:FICS}\tabularnewline
\hline 
 EBS&50&15&1.15$\times10^4$&6700&50&15&54.7&Fig.\ref{fig:EBS}\tabularnewline
\hline 
SYCS &50&50&3.8$\times10^4$&74&50&50&99.99&Fig.\ref{fig:EBS}\tabularnewline
\hline 
RDCS&50&150&1.15$\times10^5$&-9676&50&150&173&Fig.\ref{fig:EBS}\tabularnewline
\hline

\end{tabular}

\end{table*}

\begin{figure*}
	\centering
	\includegraphics[width=8 cm]{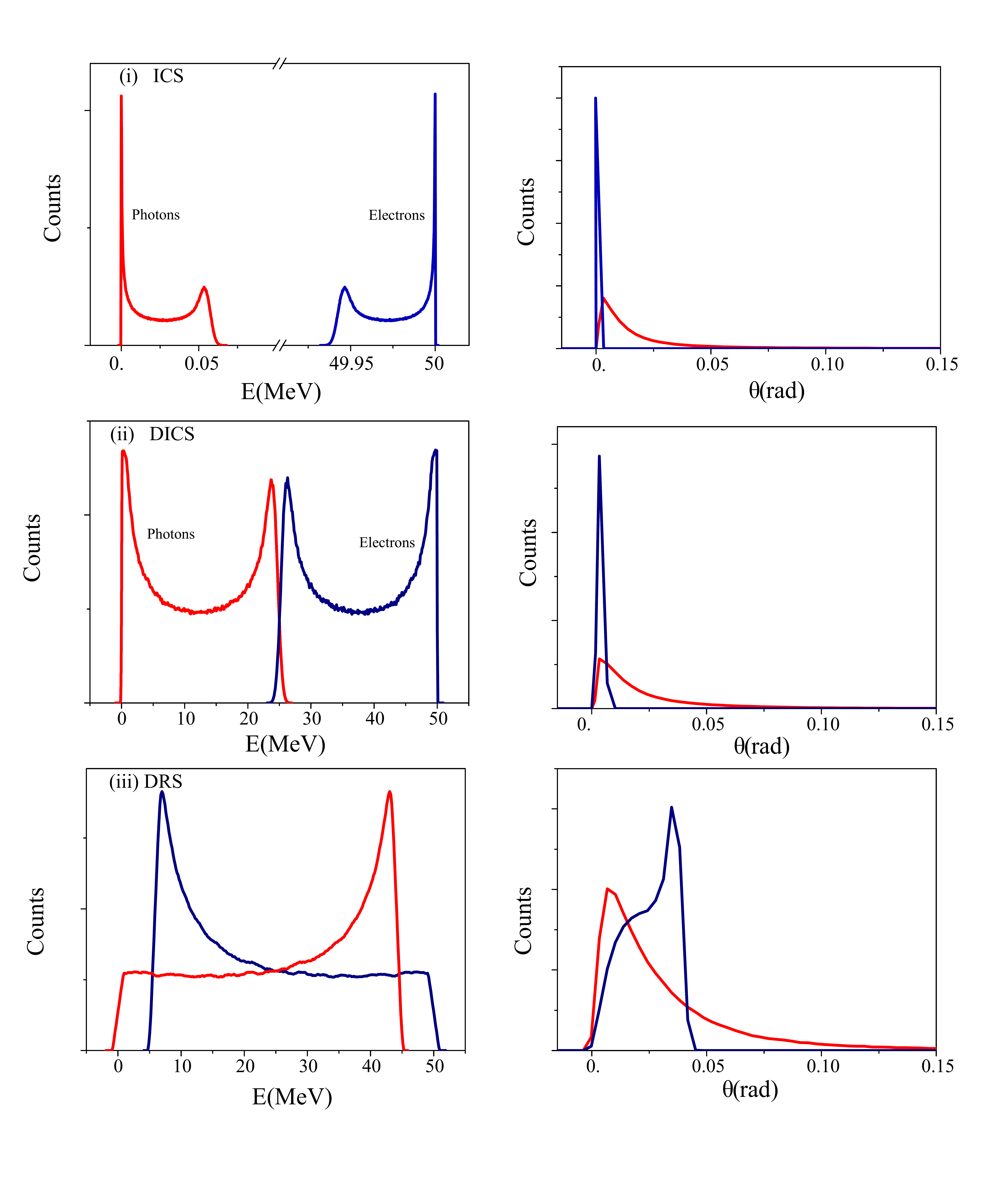}
	\caption{(i) Inverse Thomson Scattering (ITS). Left: energy distribution, right: angular distribution. Red: scattered photons, blue:scattered electrons. $E_{ph}$=1.5 eV, $E_e$=50 MeV, $bw_{ph}=5\%$. (ii) Democratic Inverse Compton Scattering (DICS). $E_{ph}$=1.3 keV, $E_e=50 MeV$, $bw_{ph}=5\%$. (iii) Deep Recoil Scattering (DRS). $E_{ph}$=10 keV, $E_e$=50 MeV, $bw_{ph}=5\%$.}
	\label{fig:ICS}
\end{figure*}

This is the region denominated "extremely relativistic" in Ref. \cite{Primakoff} and \cite{Arutyunian}, whose authors discuss the amount of energy lost by the electron in a single collision: the electron looses a large fraction of its kinetic energy in favour of the photon when collisions take place in regions (b) and (c). The way Inverse Compton Scattering in deep recoil regime can be realized is double-fold: either considering electrons with increasing initial energy $E_e$ before scattering, as in ref. \cite{Primakoff, Arutyunian} (dealing with astrophysical phenomena where incident photons are in the microwave or infrared or optical range), or increasing the energy of the incident photon, as discussed here. What matters in order to keep $X$ large is the product of the incident photon energy and the incoming electron energy: just as a side comment we note that the maximum value achieved in experiments operated in laboratories with accelerators, has been about $X=1.9$ so far (\cite{Bamber}, reporting the data of experiment E144 at SLAC) using about 45 GeV electron beams colliding with IR (1.2 eV photons) and green (2.4 eV photons) laser pulses - DICS has been observed in single photon linear Compton scattering in this experiment, as clearly reported in the photon spectra published in the paper.  There is however a special transition point, denominated Full Inverse Compton Scattering (point (2) in the Fig.s \ref{fig:regimi}-\ref{fig:50_regimi}), that is characterized by a unique absolute value of the incident photon energy, i.e. $E_{ph} = mc^2/2$, that is the divide between electrons emerging after scattering with still positive momentum (region (a) and (b)) and electrons back-scattered with negative momentum (region (c) and (d)) by the collision with incident photons. At the FICS transition point, the electron is taken down to rest in the laboratory system by the interaction with a 255.5 keV ($mc^2/2$) photon: its energy after scattering is just $mc^2$. Its kinetic energy is totally transferred to the back-scattered photon, that will propagate back towards positive z-axis direction with energy $E'_{ph} = E_e - mc^2/2 = T_e + mc^2/2$. Interesting to note that $mc^2/2$ is a value playing an important role also in direct Compton scattering, with a dual behavior than FICS: as a matter of fact if we consider an incident photon on a target, as predicted by Eq.4, with an energy much larger than the electron rest mass energy, i.e. $E_{ph} >> mc^2$, then the back-scattered photon at $\theta=0 $ will have an energy equal to $E'_{ph} = mc^2/2$ independently on the energy $E_{ph}$ of the incident photon. At FICS point the recoil factor is $X=2\gamma$. We believe that FICS can play an important role in astro-physical gamma-ray sources, where the flux of 255.5 keV X-rays can be intense and can interact with very high energy electrons generating in turns very high energetic photons in a single collision with extremely large energy transfer. At the end, FICS is the Compton scattering modality maximizing the energy/momentum transfer from the electron to the photon: this characteristic was not specifically underlined in previous literature. The impressive characteristic of FICS interaction point is the capability of 255.5 keV photons to stop "any" relativistic electron of whatsoever energy.  
Fig. \ref{fig:50_regimi} shows the different emission regimes for $E_e= 50 MeV$. $E'_{e}$ (in blue),$ E'_{ph}$ (in red) and $E_{ph}$ (in black) in MeV are represented versus $E_{ph}$ in MeV.  The stars represent few typical values chosen for spectral simulations. In Table \ref{tab:2} the values assumed by some relevant  quantities are reported.
The first of these points falls in the Inverse Thomson Scattering regime (a 50 MeV electron beam collides with a optical radiation pulse of 1.5 eV energy) and is an example of the various Compton sources operating worldwide \cite{Petrillo}. Characterized by a small recoil $X=1.1\times 10^{-3}$, this working point generates hard X-rays of edge energy $E'_{ph}$=57.4 keV, in a range useful for medical imaging applications.
The radiation spectrum, shown in Fig. \ref{fig:ICS}, window (i) in red, has the typical  bowl shape, while the electrons (in blue) lose in the interaction a negligible amount of energy. The emitted photons are all in the $1/\gamma$ cone, while the electrons maintain their straight trajectories.

\begin{figure*}
	\centering
	\includegraphics[width=15 cm]{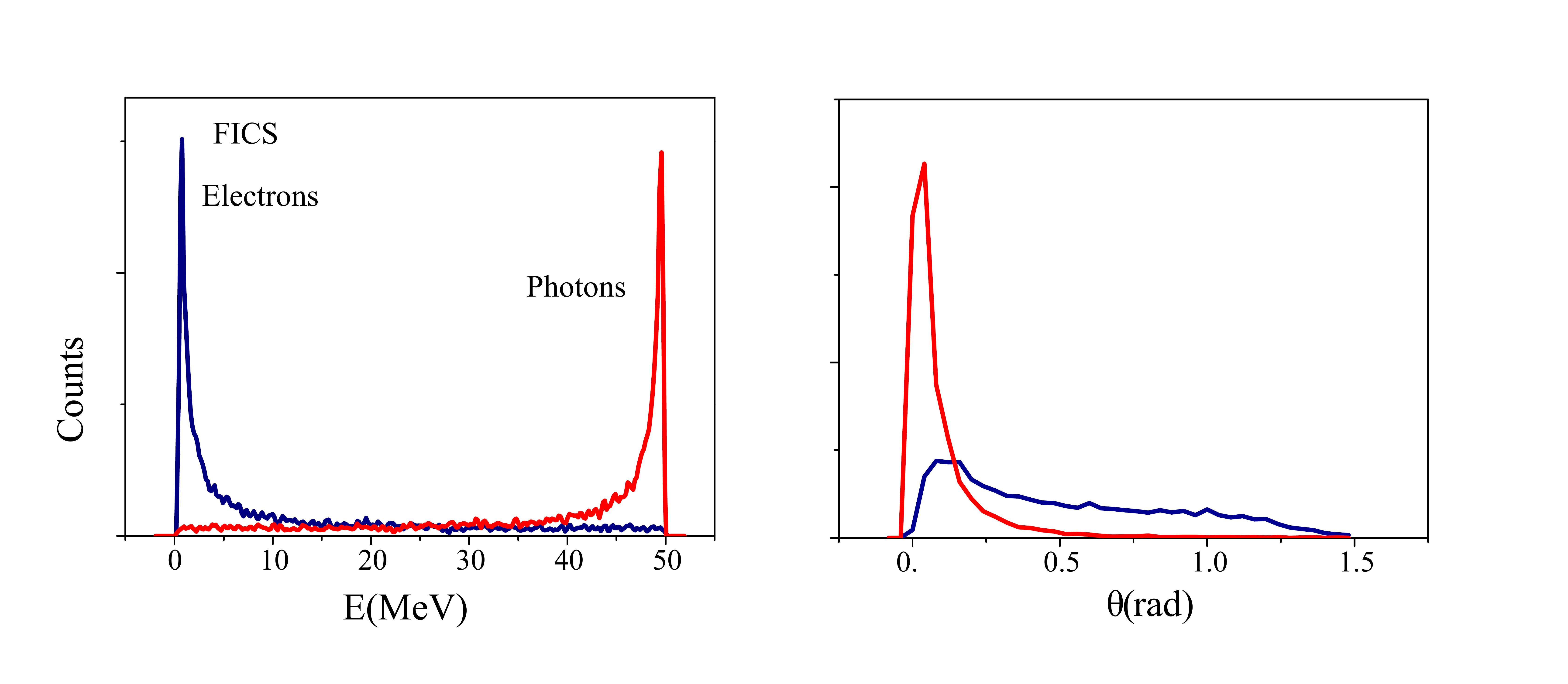}
	\caption{Full Inverse Compton Scattering (FICS). Left: energy distribution, right: angular distribution. Red: scattered photons, blue:scattered electrons.$E_{ph}=255.5 keV$, $E_e$=50 MeV, $bw_{ph}=5\%$}
	\label{fig:FICS}
\end{figure*}
When the recoil is $X=1$, corresponding to $E_{ph}$= 1.3 keV for $E_e$=50 MeV (Fig. \ref{fig:ICS}, window (ii)), the emitted photon and the emerging electron spectra maintain the broad bowl shape, while the Compton edge and the minimum electron energy limit coincide at about $E=E_e/2$. The photon angular distribution still covers the $1/\gamma$ aperture and the emerging electrons are all close to the axis.
\begin{figure*}
	\centering
	\includegraphics[width=8 cm]{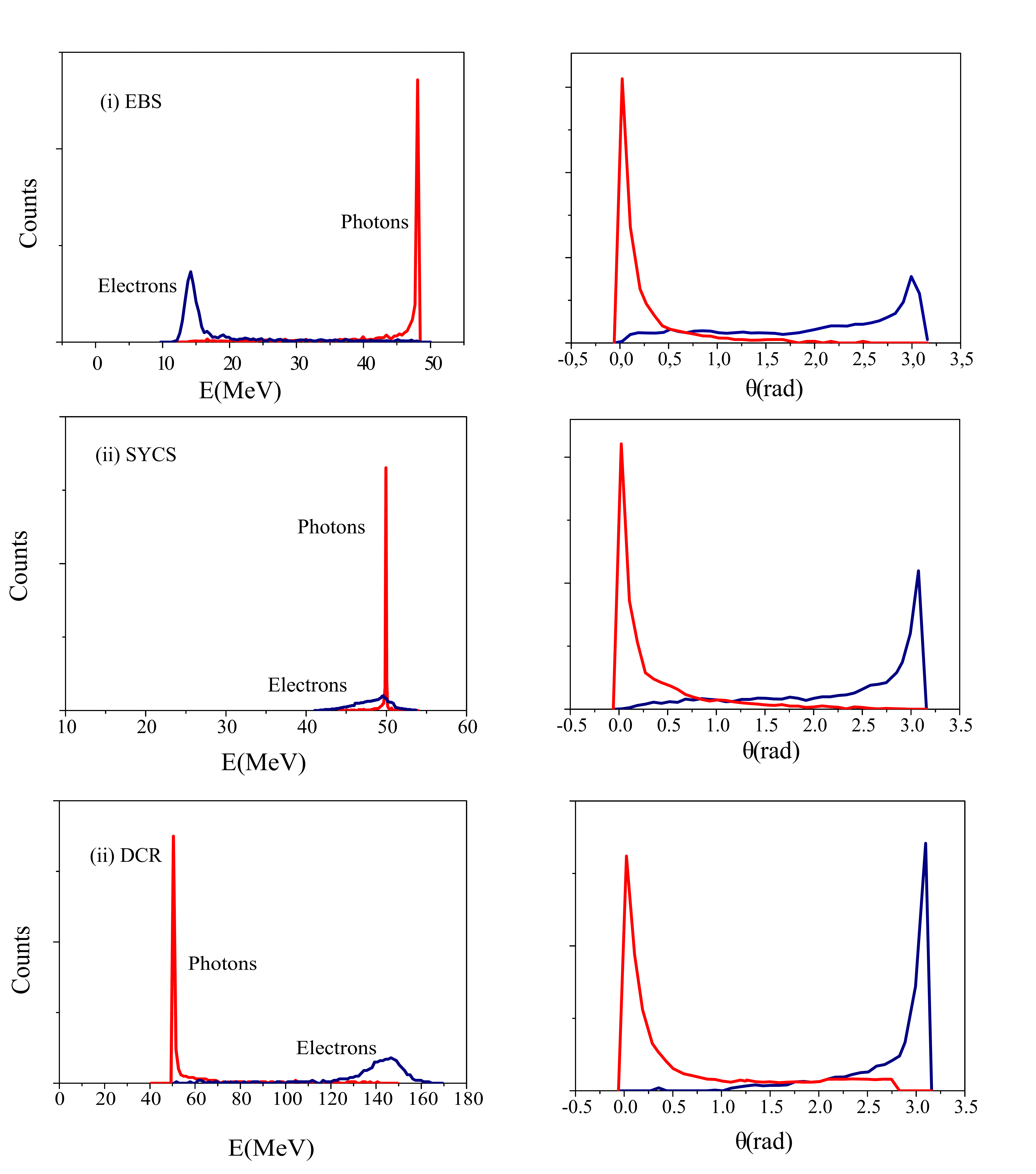}
	\caption{(i) Electron Back-Scattering (EBS). Left: energy distribution, right: angular distribution. Red: scattered photons, blue:scattered electrons. $E_{ph}=15 MeV$, $E_e=50 MeV$, $bw_{ph}=5\%$.(ii) Symmetric Compton Scattering (SYCS): $E_{ph}=50$ MeV, $E_e$=50 MeV, $bw_{ph}=5\%$. (iii) Relativistic Direct Compton Scattering: $E_{ph}$=150 MeV, $E_e$=50 MeV, $bw_{ph}=5\%$}
	\label{fig:EBS}
\end{figure*}
Increasing further the recoil, penetrating the deep recoil regime (DRS), (see Fig. \ref{fig:ICS}, window(iii)), the radiation spectrum develops a sharp forward peak, already described in Ref.s \cite{Hajima, Curatolo}. In a symmetric way, the electrons lose energy.
The angular distributions begin to enlarge.
This situation achieves its maximum expression in correspondence to the Full Inverse Compton Scattering (FICS) point, where the electron beam, hit by photons at 255.5 keV, stops whatever its energy. As shown in the simulation (Fig. \ref{fig:FICS}), there is an almost full exchange of energy between the photons and the electrons. The photon spectrum appears strongly peaked at the initial electron energy at 50 MeV, while the electrons lose almost all their kinetic energy, stabilizing at 511 keV total energy $E'_e$. While the photons still propagate in a cone around the positive z-axis, the electrons scatter in the whole solid angle and a consistent fraction revert.
Increasing the incoming photon energy beyond 255.5 keV (Fig. \ref{fig:EBS} windows (i),(ii) and (iii)), the emitted photon energy remains attested at the initial electron energy value (namely 50 MeV), the remaining energy being acquired by the electrons which now move faster and faster in the reverse direction.
The radiation spectrum shrinks, reaching the full monochromaticity in correspondence of the Symmetric Inverse Compton Scattering (SYCS) point (\ref{fig:EBS}, window (ii)), defined by the condition $E_e=E_{ph}$ and widely analyzed in Ref. \cite{Serafini} also in general to non-relativistic electrons, for which the SYCS condition actually applies to the electron and photon momentum, i.e. $p_e=\hbar k$. 
We note that a relatively broad bandwidth of the initial photon beam does not change the thinness of the final spectrum. In Fig. \ref{fig:EBS}, the initial photon relative bandwidth is assumed 5\%, while in the symmetric Compton condition the final bandwidth is under $0.2\%$. Conversely, the electrons, even if monoenergetic at the beginning, exit the collision dispersed in energy.
\begin{figure*}
	\centering
	\includegraphics[width=10 cm]{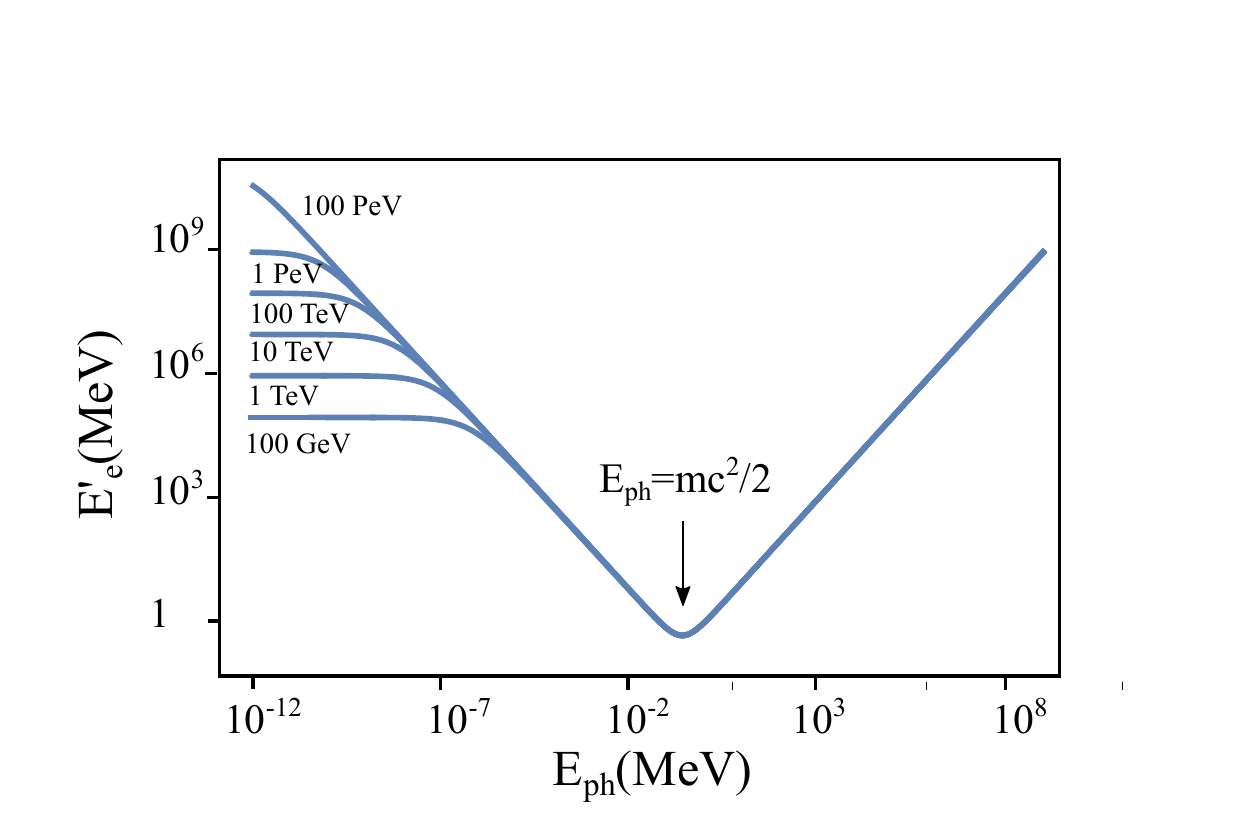}
	\caption{$E'_{ph}$ as function of $E_{ph}$ for various values of initial energy $E_e$, namely 100 GeV,1,10,100 TeV, 1,100 PeV. Note that all electrons stop after collision when $E_{ph}= mc^2/2$=0.2555 MeV, independently on their initial energy $E_e$. The incident photon energy region $E_{ph}= 10^{-10}$ to $10^{-9}$ MeV corresponds to the Cosmic Microwave Background Radiation (CMBR), while the region $E_{ph}= 10^{-7}$ to $10^{-5}$ MeV corresponds to Extra-Galactic Light (EBL) and Inter-Stellar Radiation Field (ISRF), ref. \cite{Vernetto}.}
	\label{fig:minimo}
\end{figure*}
The unbelievable properties of the FICS working point are highlighted in Fig. \ref{fig:minimo}.
Spanning a very large range of incident photon energies, from 1 $\mu eV$ up to 1 TeV (in order to cover the energy range of intra-galactic and extra-galactic photon background, spanning from CMBR to EBL and ISRF, see ref. \cite{Vernetto}), and considering interacting electron energies of 100 GeV,1, 10, 100 TeV, and 1, 100 PeV, all curves plotted in Fig.\ref{fig:minimo} show the energy of the electron after scattering $E'_e$ : this quantity exhibits a common minimum at $mc^2/2$ incident photon energy with a common value of $ E'_e = mc^2$ independent on the initial electron energy $E_e$. It is also interesting to notice that the behaviour of $E'_e$ as function of $E_{ph}$ becomes independent on the initial electron energy $E_e$ as soon as the recoil factor $X$ becomes much larger than 1, as well shown in Fig. \ref{fig:minimo}. This is clearly due to the fact that for large recoil the energy of the back-scattered photon asymptotically tends to the energy of the incident electron, therefore $E'_e$ asymptotically tends to $E_{ph}$. Nevertheless, we need to underline that increasing the energy available in the center of mass $E_{cm}$, as clearly inherent in data shown in Fig.\ref{fig:minimo}, other reactions become competitive and statistically more likely than Compton scattering, like triplet production (above $E_{cm}$= 1.5 MeV) , muon pair production (above  $E_{cm}$=210 MeV) and other hadronic reactions involving pion/meson production, that become dominant only above $E_{cm}=$ 600 MeV (see ref. \cite{Burk}), that implies $E_e= $360  GeV for 255.5 keV incident photons. A specific discussion on cross section behaviors of different QED processes in this energy range can be found in ref. \cite{EXMP}. The behavior shown in Fig.\ref{fig:minimo} is strictly relative to the kinematics of inverse Compton scattering in exotic regions of the incident particle energy range never considered previously in the literature under such specific details.
In all regimes, the number of events is proportional to the beam luminosity and weighted by the cross section \cite{KleinNishina}.
The total unpolarized Compton cross section is:
\begin{equation}\begin{split}
\sigma= \frac{2\pi r_e^2}{X}\left[\frac{1}{2}+\frac{8}{X}-\frac{1}{2(1+X)^2}+\right.\\
\left.\left(1-\frac{4}{X}-\frac{8}{X^2}\right)\log(1+X)\right]. 
\end{split}
\end{equation}
The values of $\sigma$ vary between the classical limit $X\to 0$ and the ultra-relativistic limit $X\to \infty$ as: 
\begin{equation}
\left\{
\begin{aligned}
&\lim_{X\to 0}\sigma=\frac{8\pi r_e^2}{3}(1-X)=\sigma_T(1-X) \\
&\lim_{X\to \infty}\sigma=\frac{2\pi r_e^2}{X}\left(\log{X}+\frac{1}{2}\right)
\end{aligned}
\right.
\end{equation}
where $r_e$ is the classical electron radius and $\sigma_T=0.67$ barn is the total Thomson cross section \cite{Jackson}. In particular, when the recoil factor is very large the relationship $X=(E_{cm}/mc^2)^2$ holds, so the cross section scales roughly like the inverse square of the center of mass energy, i.e.
\begin{equation}
\lim_{X\to \infty}\sigma \simeq (E_{cm}/mc^2)^2(0.5+\log(E_{cm}/mc^2)^2),    
\end{equation}
decreasing the probability to undergo Compton scattering between the incoming electron and the incident photon. Nevertheless, as extensively discussed in ref. \cite{Serafini}, the angular cross section peaks at $ \theta=0$, implying that the most likely scattering is back-scattering of the photon, condition of maximum energy/momentum exchange between the electron and the photon, at which the values reported in Table I correspond. This may be the peculiarity of Compton scattering w.r.t. other two-particle interaction, like for example Bhabha scattering or Moeller scattering, as is well shown in the evolution of the spectral distribution visible from Fig.\ref{fig:ICS} to Fig. \ref{fig:EBS}.

\section{Deep recoil and Symmetric Compton Scattering cancel all angular correlations}

All radiation emission by relativistic charged particles that involves a blue-shift based on relativistic Doppler effect is characterized by an angular correlation of the emitted photon energy vs. the angle of emission w.r.t. the propagation axis of the emitting particle. This is represented in all formulas by the well know $\gamma^2\theta^2$ term present in synchrotron radiation, ICS, betatron radiation, etc. Photons emitted at a non-zero $\theta$ angle are red-shifted w.r.t. that emitted at $\theta=0$. On the other side the radiation emission is concentrated within a $\theta=1/\gamma$ cone. It is only SYCS that cancels completely such an angular correlation: we rewrite Eq.3 by expressing the formula only in terms of the recoil parameter $X$, the asymmetry factor $A$ and the Lorentz factor $\gamma$. $A$ is generally defined, for any value of the electron velocity $\beta$ and Lorentz factor $\gamma$ as $A=\beta\gamma^2-X/4$. We obtain
\begin {equation}
E'_{ph}=\frac{\gamma^2+A+X/4}{\gamma^2-A cos\theta+X/4}E_{ph}
\end{equation}

that is at all equivalent to Eq. 3. Stating that at the SYCS transition point ($A=0$) the dependence of the back-scattered photon energy $E'_{ph}$ on the scattering angle $\theta$ is cancelled (see ref. \cite{Serafini} for further discussion).
While SYCS truly cancels the angular correlation with the scattering angle, any large recoil regime $X\gg1$ is characterized by a strong damping of the angular correlation $\gamma^2\theta^2$. We can re-write Eq.3 by assuming both $\gamma>>1$ and $\theta<1/\gamma$ (see Ref. \cite{Curatolo}) as

\begin {equation}
E'_{ph}=\frac{4\gamma^2E_{ph}}{1+\gamma^2\theta^2+X}
\end{equation}

that can be asymptotically expressed for large values of $X\gg1$ as

\begin {equation}
E'_{ph} \simeq E_e (1-\frac{1+\gamma^2\theta^2}{X})
\end{equation}

clearly showing the damping of angular correlation at large recoils. As extensively discussed in ref. \cite{Curatolo} and \cite{Serafini}, such a damping applies also to the angular spread at collision due to the electron beam emittance, making a deep recoil ICS insensitive to beam emittance, a great advantage in designing ICS sources for very mono-chromatic X and $\gamma$ rays.

Another interesting effect of the large recoil regime is the cancellation of the dependence of back-scattered photon energy on the collision angle. This is another quite surprising effect, that can be appreciated by recalling the general formula (see, for instance, \cite{twocolor})
\begin{equation}
E'_{ph}=E_{ph} \frac{\gamma^2 (1-\beta cos\alpha)}{\gamma^2 (1-\beta cos\theta)+\frac{X}{4}(1-cos(\alpha+\theta))}
\end{equation}
that reduces to Eq.3 when the collision angle $ \alpha$ is equal to $\alpha=\pi$, i.e. at head-on collision.

When the recoil factor is very large the previous formula asymptotically tends to

\begin {equation}
E'_{ph}\simeq\frac{4\gamma^2E_{ph}(1-cos\alpha)/2}{1+\gamma^2\theta^2+X(1-cos\alpha)/2}
\end{equation}

Considering photons fully back-scattered at $\theta=0$ previous equation reduces to

\begin {equation}
E'_{ph}\simeq E_e(1-\frac{2}{X(1-cos\alpha)})
\end{equation}

clearly showing the de-sensitivity to collision angle $\alpha$ induced by large recoils. This peculiar property opens the way to designing sources of mono-chromatic gamma-rays based on the interaction of mono-energetic low energy electron beams with photon hohlraums, where large X-ray photon densities are stored (see ref. \cite{Pike}). If the electron-photon collision inside the hohlraum is taking place at large recoil values, then the energy of the secondary gamma-ray beam generated (leaving the hohlraum in the same direction of propagation of the electron beam) does not depend on the large dispersion of the collision angle inside the thermal X-ray photon bath. The dependence on the thermal photon energy spread is also cancelled by large recoil interactions, through a spectral purification mechanism, as discussed in next section.

\section{Spectral purification induced by deep recoil}
\begin{figure*}
	\centering
	\includegraphics[width=18 cm]{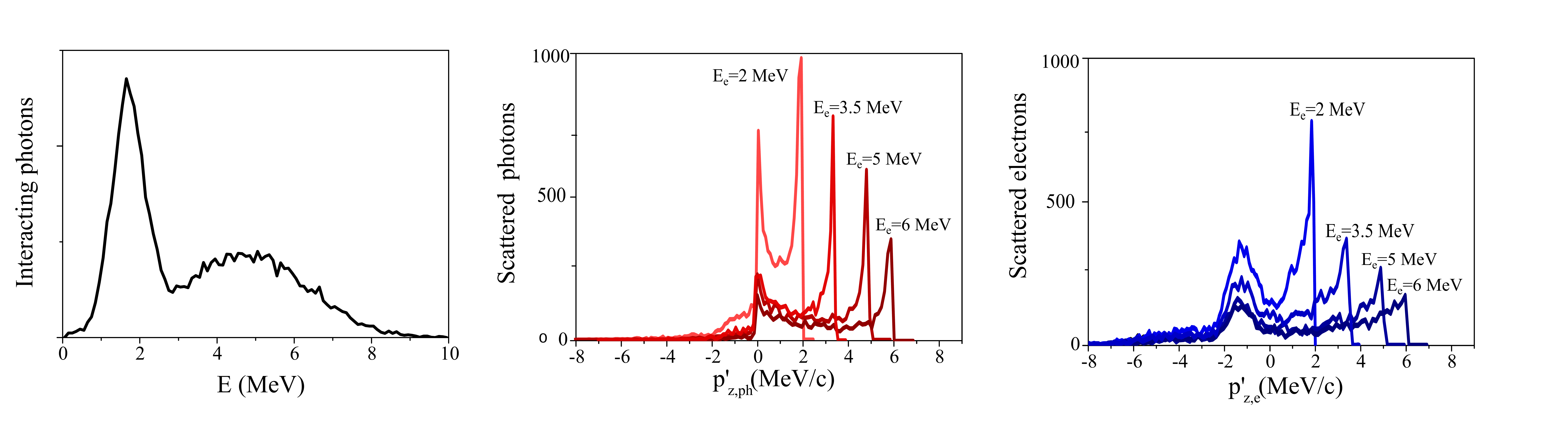}
	\caption{Left window: Spectrum of the incident photon similar to channelling radiation in crystals, with colliding electron beams of energy $E_e=2 MeV, 3.5 MeV, 5 MeV, $ and $ 6 MeV$. Central window: Momentum spectrum of scattered photons. Right window: Momentum spectrum of scattered electrons}
	\label{fig:Channel}
\end{figure*}\begin{figure*}
	\centering
	\includegraphics[width=10 cm]{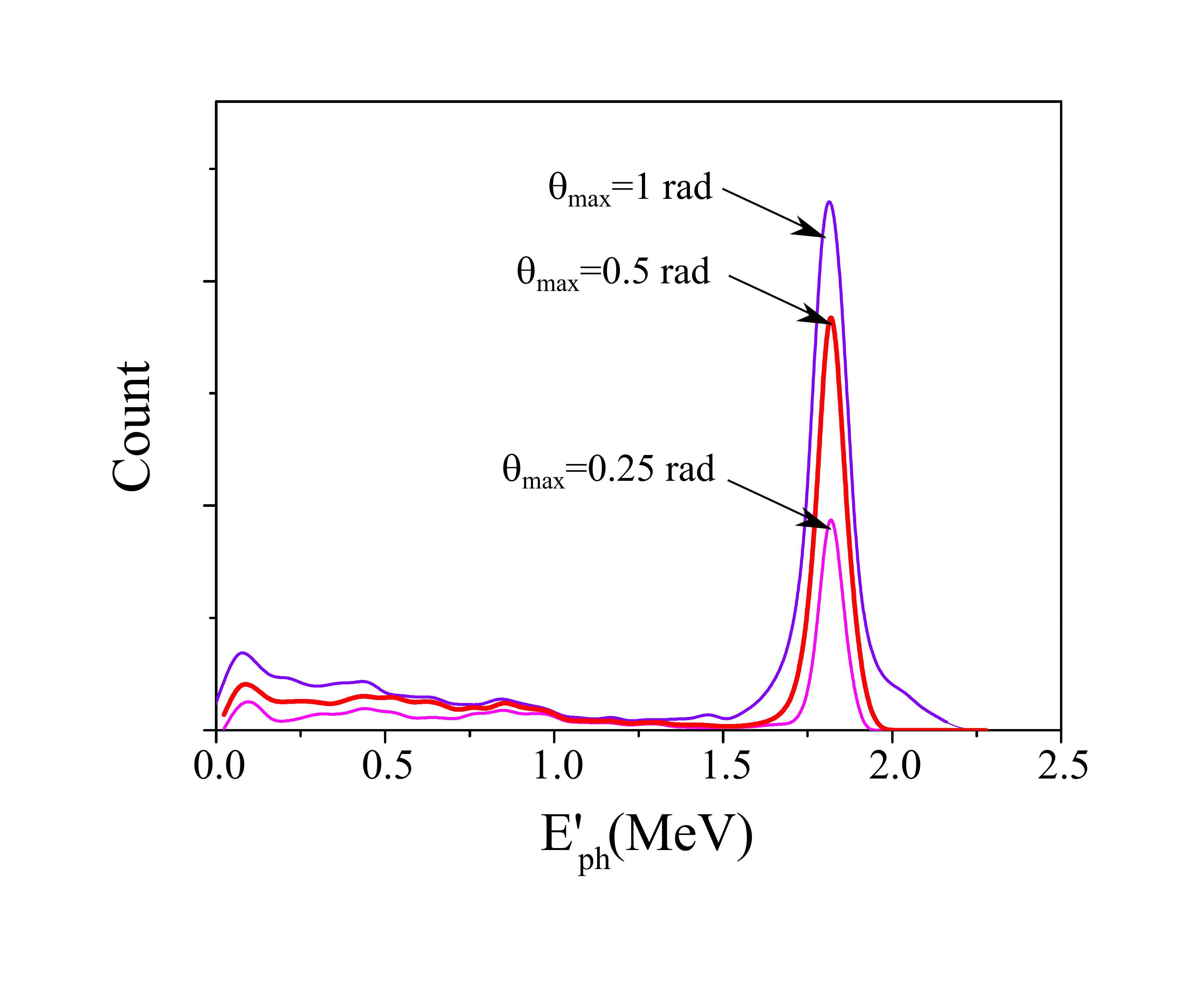}
	\caption{Collimated energy spectrum. $E_e=1.9 MeV$, curve magenta:$\theta_{max}= 0.25 rad$, red:$\theta_{max}= 0.5 rad$,blue: $\theta_{max}= 1 rad$}
	\label{fig:puri}
\end{figure*}

The peculiar spectral behavior around the SYCS point can be exploited to conceive an extremely mono-chromatic source of $\gamma$ rays. Starting from broadband radiation sources ($\Delta E_{ph}/E_{ph}\simeq 30\%$), as channeling radiation \cite{Bandiera} or betatron emission from plasma accelerated beams \cite{EupraXia}, a highly monochromatic gamma-ray beam can be produced by colliding these photons with electron beams of similar energy .

Once again large recoils reduce the dependence of the scattered photon beam bandwidth on the bandwidth of the incident photon beam, similarly to what happens with angular correlations (as discussed in previous section). Recalling that for large recoil factors $X\gg1$ the asymptotic expression of the back-scattered photon at $\theta=0$ is given by $E'_{ph}\simeq (1-1/X)E_e$ ,  we can easily derive the relative bandwidth of the back-scattered photon beam $\Delta E'_{ph}/E'_{ph}$  as:

\begin {equation}
\frac{\Delta E'_{ph}}{E'_{ph}} \simeq \frac{1}{X} \frac{\Delta E_{ph}}{E_{ph}}
\end{equation}

as a function of the incident photon beam relative bandwidth  $\Delta E_{ph}/E_{ph}$. In agreement with ref. \cite{Curatolo} as far as the condition $X\gg1$ is assumed.

This damping of the incident photon beam bandwidth, that is compressed by a factor $1/X$ into the back-scattered photon beam bandwidth, is very clearly illustrated in Fig. \ref{fig:Channel}, showing the initial energy spectrum (left window) and the emitted photon (central window) and electron (right window) momentum spectra, for the case of an incident broad-band photon beam generated by channeling radiation in a crystal (ref. \cite{Bandiera}) and a colliding mono-energetic electron beam with variable energy, spanning the 2 - 6 MeV energy range. The photon beam spectra with a highly monochromatic peak shown in the central diagram clearly demonstrate the spectral purification mechanism enabled by the deep recoil inverse Compton scattering regime. Here what matters to achieve spectral purification is just a large value of the recoil factor $X$, which vary in a range from 64 (minimum value at 2 MeV colliding electron beam) up to about 200 (maximum value in case of 6 MeV electrons). The very peculiar point is that this example of beam collision spans several regions of ICS, since the electron beam Lorentz factor $\gamma$ varies from about 5 up to nearly 12, implying that the 2 MeV case is fully in RDCS regime, while 3.5 MeV is spanning across RDCS, SYCS and EBS, 5 MeV covers EBS and FICS and 6 MeV FICS and DRCS. As a consequence, the spectrum of the electron momentum (Fig. \ref{fig:Channel}, right window) resembles the initial photon spectrum, but at the turn of the zero with an electron ensemble going forward and another going backward. No matter what regime is in action, as far as the recoil factor $X$ is a large number, high spectral purification applies. Fig. \ref{fig:puri} presents instead the collimated energy spectrum, taken by selecting the radiation inside the angles $\theta_{max}$=0.25 rad$ (\simeq 1/\gamma)$, $\theta_{max}$=0.5 rad $\theta_{max}$=1 rad, where the spectral purification manifests itself evidently. Let us also underline the great advantage of using these deep recoil regimes to generate highly mono-chromatic gamma-rays: the energy of the colliding electron beam needed to generate gamma-ray beams in the MeV energy range is not much larger than a few MeV. Unlike typical ICS sources aimed at gamma-ray beam generation in the ITS regime \cite{ELI} where GeV-class electron beams are needed to enable applications in the photo-nuclear physics and photonics. Compactness and sustainability are certainly in favour of the deep recoil regime.

\section*{}
\section{Conclusions} We revisited the kinematics of Compton Scattering, the electron-photon interactions producing electrons and photons in the exit channel, covering the full range of energy/momenta distribution between the two colliding particles, with a dedicated view to statistical properties of secondary beams that are generated in beam-beam collisions. Starting from the Thomson inverse scattering, where electrons do not recoil and photons are back-scattered to higher energies by a Lorentz boost effect (factor $4\gamma^2$), we analyze three transition points, separating four regions. These are in sequence, given by increasing the electron recoil (numbers are for transition points, letters for regions): a) Thomson back-scattering, with a recoil factor $X \ll 1 $ 1) equal sharing of total energy in the exit channel between electron and photon (recoil factor X=1, b) deep recoil regime where the bandwidth/energy spread of the two interacting beams are exchanged in the exit channel (here $1\ll X \ll 2\gamma$) 2) electron is stopped, i.e. taken down at rest in the laboratory system by colliding with an incident photon of $mc^2/2$ energy and the recoil factor is $X=2\gamma$, c) electron back-scattering region, where incident electron is back-scattered by the incident photon, with a recoil $2\gamma<X<4\gamma^2$ 3) symmetric scattering, when the incident particles carry equal and opposite momenta, so that in the exit channel they are back-scattered with same energy/momenta and $X=4\gamma^2$, d) Compton scattering ($a'$ $la$ Arthur Compton, see ref.4), where photons carry an energy much larger than the colliding electron energy, occurring when $X>4\gamma^2$. For each region and/or transition point we discussed the potential effects of interest in diverse areas, like generating mono-chromatic gamma ray beams in deep recoil regions with spectral purification, or possible mechanisms of generation and propagation of very high energy photons in the cosmological domain.

Acknowledgements: we acknowledge useful discussions with Laura Bandiera and Gianfranco Paterno'.

\end{document}